\documentclass[twocolumn,a4paper,preprintnumbers,amsmath,amssymb,nofootinbib,floatfix]{revtex4}
\usepackage{hyperref}
\usepackage{cleveref}
\usepackage{color}
\usepackage{graphicx}
\usepackage{dcolumn}
\usepackage{bm}
\usepackage{latexsym}
\usepackage{epsfig}
\usepackage{rotating}
\usepackage{bm}

\newcommand{\be}{\begin{equation}}
\newcommand{\ee}{\end{equation}}
\newcommand{\beqq}{\setlength\arraycolsep{2pt}\begin{eqnarray}}
\newcommand{\eeqq}{\vspace{0cm} \end{eqnarray}}
\newcommand{\bea}{\begin{eqnarray}}
\newcommand{\eea}{\end{eqnarray}}

\newcommand{\lambdab}{\stackrel{\neg}{\lambda}}
\newcommand{\xib}{\stackrel{\neg}{\xi}}

\begin{document}

\title{From inflation to recent cosmic acceleration: The fermionic Elko field driving the evolution of the universe}

\author{S. H. Pereira$^{1}$} \email{shpereira@feg.unesp.br}
\author{T. M. Guimar\~aes$^{1,2}$} \email{thiago.mogui@gmail.com}

\affiliation{$^{1}$Universidade Estadual Paulista (Unesp)\\Faculdade de Engenharia, Guaratinguet\'a \\ Departamento de F\'isica e Qu\'imica\\ Av. Dr. Ariberto Pereira da Cunha 333\\
12516-410 -- Guaratinguet\'a, SP, Brazil\\
$^{2}$Instituto Federal do Paran\'a - Campus Ivaipor\~a\\
86870-000 - Ivaipor\~a - PR - Brazil}




\begin{abstract}
In this paper we construct the complete evolution of the universe driven by the mass dimension one dark spinor called Elko, starting with inflation, passing by the matter dominated era and finishing with the recent accelerated expansion. The dynamic of the fermionic Elko field with a symmetry breaking type potential can reproduce all phases of the universe in a natural and elegant way. The dynamical equations in general case and slow roll conditions in the limit $H\ll m_{pl}$ are also presented for the Elko system. {Numerical analysis for the number of e-foldings during inflation, energy density after inflation and for present time and also the actual size of the universe are in good agreement with the standard model of cosmology. An interpretation of the inflationary phase as a result of Pauli exclusion principle is also possible if the Elko field is treated as an average value of its quantum analogue.}
\end{abstract}

\maketitle


\section{Introduction}

A model that correctly describes the whole evolution of the universe is one of the main challenge of modern cosmology. In the standard model, the universe starts with the big bang in a very hot and dense phase dominated by quantum effects, while the energy density is greater than Planck energy. Then the universe suffer an abrupt expansion known as \emph{inflation} and evolves dominated by its material content, first radiation, followed by dark matter and finally some kind of dark energy at late times. All models describing these phases can be tested with great precision after the recent results from Planck mission \cite{Planck2013}. In particular, dozens of inflationary models based on a single-scalar field have been considered in \cite{Encinflat1,Encinflat2} by using Bayesian statistical analysis and surprisingly they found that  the better model is described by the simplest version of inflation, a kind of vacuum decay model. A natural question that appears is: what happens to the scalar field after inflation, since the universe must enter a matter dominated phase? The answer coming from standard model is that the universe passes to a reheating phase, where the oscillation of the scalar field transfers energy to radiation and then to matter dominated era. {In this way, any model that would be an alternative to the evolution of the universe must describe all these phases satisfactorily.}

Very recently, a class of mass dimension one fermions named Elko, initially proposed by Ahluwalia and Grumiller \cite{AHL1,AHL2,ahl2010a,ahl2010b,ahl2011a,AHL3,AHL4} as a natural candidate to a fermionic dark matter particle, underwent a profound overhaul in the definition of their duals, culminating with a field that is local and Lorentz covariant \cite{AHL4,WT} (see also \cite{RJ} for some additional support). Thus, the interest for such class of non-standard spinors (NSS), or dark spniors, has increased in recent years, since they are naturally neutral and has mass dimension one\footnote{Dirac fermions have mass dimension $3/2$}, which leads them to satisfy only a Klein-Gordon type equation. The Elko field is constructed as a spin-$1/2$ field describing fermions that are eigenstate of the charge conjugation operator\footnote{Dirac fermions are eigenstate of parity conjugation operator}. Moreover, as neutral fields, they are good and natural candidate to particles of dark matter in the universe, an open problem in cosmology. Models in which the Elko field is considered as candidate to dark matter or dark energy in the universe have been proposed recently \cite{BOE1,BOE2,BOE5,BOE3,BOE4,FABBRI,FABBRI1,BOE6,GREDAT,
BASAK,BOE7,WEI,basak,sadja,kouwn,sajf,saj,js,sa,asf}. In the present work we are interested in the Elko field as a possible candidate to drive inflation, following a matter dominated era and finishing as the responsible for the recent cosmic acceleration of the universe. In this final phase, the Elko field just rolls down to the minimum of a potential and acts as a cosmological constant term. 

{In order to compare the equations and results from a single scalar field driving the inflation with the corresponding ones of Elko field, we present here a brief review of the scalar field in a flat Friedmann-Robertson-Walker (FRW) background, following the references \cite{Encinflat1,Encinflat2,bookliddle,linde1,lindePLB}.}

The Einstein equations for a single scalar field $\phi$ in a flat FRW metric are:
\begin{equation}
H^2=\frac{\kappa^2}{3}[\frac{1}{2}\dot{\phi}^2 + V(\phi)]\,,\label{eqH}
\end{equation}
\begin{equation}
\dot{H}=-\frac{\kappa^2}{2}\dot{\phi}^2\,,\label{eqHdot}
\end{equation}
\begin{equation}
\ddot{\phi}+3H\dot{\phi}+V'(\phi)=0\,,\label{eqphi}
\end{equation}
with $V'(\phi)\equiv dV/d\phi$, $H\equiv\dot{a}/a$ the Hubble expansion parameter, $\kappa^2\equiv8\pi G=1/m^2_{pl}$ with $c=1$ and $m_{pl}\approx 10^{19}$GeV is the Planck mass. 

The energy density and pressure for the scalar field are given by 
\begin{equation}
\rho={\dot{\phi}^2\over 2}+V(\phi)\,,\label{rho}
\end{equation}
\begin{equation}
p={\dot{\phi}^2\over 2}-V(\phi)\,.\label{pphi}
\end{equation}
Given a potential $V(\phi)$,  inflation occurs if the slow-roll parameters $\epsilon$ and $\eta$ satisfies \cite{bookliddle}: 
\begin{equation}
\epsilon(\phi)\equiv \frac{\vert\dot{H}\vert}{H^2}\simeq \frac{1}{\kappa^2}\bigg(\frac{V'(\phi)}{V(\phi)}\bigg)^2 \ll 1\,,\label{eqVl}
\end{equation} 
\begin{equation}
\vert\eta(\phi)\vert \equiv \Bigg\vert\frac{\ddot{\phi}}{H\dot{\phi}}\Bigg\vert \simeq \Bigg\vert\frac{1}{\kappa^2}\frac{V''(\phi)}{V(\phi)}\Bigg\vert \ll 1\,,\label{eqVll}
\end{equation}
which justifies to neglect the kinetic term from (\ref{eqH}) ($\dot{\phi}^2/2\ll V(\phi)$) and the acceleration term from (\ref{eqphi}) ($\ddot{\phi}\ll 3H\dot{\phi}$).  Although being necessary conditions to drive inflation, the smallness of such parameters is not sufficient to guarantee that those terms can be neglected \cite{bookliddle}. Sometimes, the additional assumption $\dot{\phi} \simeq -V'/3H$ is also needed. Such parameters are used to restrict the form of possible potentials and these conditions ensure that the onset of expansion is approximately exponential, as required by all inflationary theories.

An alternative expression for the condition to inflation occur is given by \cite{bookliddle}:
\begin{equation}
\frac{d}{dt}\frac{H^{-1}}{a}<0\,,\label{condinflation}
\end{equation}
showing that the comoving Hubble length $H^{-1}/a$ is decreasing with time. 

Dozens of potential have been proposed in last decades in order to drive the inflationary phase of expansion of the universe. Some of them have physical motivations whilst other are just placed by hand in order to furnish correct results. {References \cite{Encinflat1,Encinflat2} makes a detailed statistical analysis on several potentials based on recent Planck mission results \cite{Planck2013}, indicating which of them are good or not to drive inflation. We just cite some potential of interest, as power law potential (or chaotic inflation), exponential potential, inverse power law potential, hill-top models, symmetry breaking potential, natural inflation, hybrid inflation (or multi fields inflation), among others. In all these potentials there are parameters to be adjusted according to duration of inflation for instance, measured by the number of e-foldings $N\equiv\ln(a_{end}/a_{initial})$. Also to measure the correct transition to the end of the inflationary era, called reheating phase and most important, the correct prediction of density perturbations, responsible for the formation of galaxies and cluster of galaxies and also for the anisotropies in the cosmic microwave background (CMB) radiation. This last test throw away several potentials based on Planck 2013 observations \cite{Planck2013}.}

As already pointed out above, another important characteristic of inflationary models is how inflation ends. The scalar field, after rolling down to the bottom of the potential, needs to leave the scene in order to next phase of the universe takes place. In other words, the scalar field must decay to its minimum value in order not to act any more. This process is called reheating. A hot universe at the end of the inflation is a necessary condition in order to radiation dominate and also the conventional matter start to form while the temperature is cooling down. In the modern inflationary model, the scalar field oscillates while rolls down to the bottom of the potential, transferring energy to other matter fields, or even decaying into standard particles. The details of reheating are an important subject into the inflationary cosmology. Some models add a phenomenological decay term to the equation of motion of the scalar field (after inflation, the term $\ddot{\phi}$ is important again):
\begin{equation}
\ddot{\phi}+3H\dot{\phi}+V'(\phi)+\Gamma\dot{\phi}=0\,,\label{eqphiG}
\end{equation}
where $\Gamma$ is considered a decay rate of the field $\phi$ into other particles \cite{bookliddle,kolb,linde2}. Such friction term is also needed to make fine adjustments in theory, not allowing inflation to occur forever for instance. {Such term can be calculated in standard models of inflation in
terms of the coupling between the inflaton field and the particles it couples in order to control the damping of oscillations during the reheating \cite{linde2}}.

After all, having the inflation occurred, the scalar field rolling down to the bottom of the potential and oscillating accordingly with the last term of (\ref{eqphiG}) to correctly stop the inflation, a final key question still prevails. What is such scalar field? It is named inflaton, the particle responsible for the inflation, but his very nature is not known yet. {The only fundamental scalar particle detected in the nature is the Higgs field, responsible for the electro-weak symmetry breaking, which also put scalar field
based inflation scenarios as the most attractive ones. For this reason alternative models would also have the same behaviour as the standard scalar field one.}

All the observed matter in the universe are constituted by fermions, which motivated us to construct an inflationary model with the fermionic dark spinor called Elko. {Being a good candidate to describe dark matter in the universe we have found that it can drive the inflation and the dark matter evolution after inflation, very close to standard scalar field based models. Also, the recent accelerated expansion of the universe can be correctly described if a cosmological constant like term be added to the potential of the field.} Finally, since the dark matter does not interact electromagnetically with other baryonic matter, the presence of {an additional electromagnetic radiation term into the original Lagrangian would reproduce the radiation dominated phase after inflation, responsible for the nucleosynthesis}, before the dark matter dominance\footnote{We recommend to the reader the books by Kolb-Turner \cite{kolb}, specifically pages 73 and 274 where the thermal history of the universe is traced by interesting figures, with very realistic orders of magnitude for some parameters of the standard  model of cosmology, both during and after inflation. Also the most up-to-date books by Weinberg \cite{weinberg} and Peter \& Uzan \cite{uzan} are good references.}. {In the present study we do not include an electromagnetic radiation term, thus only inflation, dark matter dominance and accelerated expansion are addressed in this model. We aim to place the Elko inflationary model as an alternative to standard scalar fields based models once it has basically the same behaviour but could also describe the missing dark matter in the universe, which we believe to be described by a fermionic field.}

{The paper is organized as follows. In Section II, we introduce the main Elko equations in cosmology. In Section III we study numerically how Elko field can be a good candidate to drive inflation, dark matter and late time cosmic acceleration. Section IV finish with some concluding remarks. Appendix A include a brief deduction of the main equations used in Section II, for completeness. Appendix B present some details on the numerical analysis of the coupled system of equations.}

\section{Elko dynamics in FRW}

{In momentum space, the mass dimension one fermionic Elko field \cite{AHL1,AHL2,ahl2010a,ahl2010b,ahl2011a,AHL3,AHL4} is represented by $\lambda_{\beta}({\bf k})^{S/A}$ and constructed as fermions that are eigenstate of the charge conjugation operator $C$, satisfying a relation of type $\stackrel{\neg}{\lambda}_{\beta}({\bf k})^{S/A}\lambda_{\beta^{'}}({\bf k})^{S/A} = \pm 2m\delta_{\beta\beta^{'}}$, where $\lambda_{\beta^{'}}({\bf k})^{S/A}$ and $\stackrel{\neg}{\lambda}_{\beta}({\bf k})^{S/A}$ are the usual spinor and its dual, respectively, the index $\beta$ stands for the two possible helicities of the spinor and $S/A$ stands for the self-conjugate spinor ($S$) and anti-self-conjugate ($A$). {The dual must be conveniently defined, and the first formulation defined it as $\stackrel{\neg}{\lambda}_{\beta} =i \varepsilon^\alpha_\beta \lambda_{\alpha}^\dagger\gamma^0$, with $\varepsilon^\alpha_\beta = - \varepsilon^\beta_\alpha = +1$ and $\gamma^0$ the Pauli matrix. More details can be found in \cite{AHL4,RJ,WT} with another definition for the dual}. While profoundly altering the quantum structure of the field, the definition for the dual of Elko does not alter its classical formulation, so that cosmological applications remain valid. The four spinors $\lambda_{\beta}({\bf k})^{S/A}$ will act as expansion coefficients to construct the quantum analogue of the field, which we will call just as $\lambda(x^\mu)$. The positivity of energy will requires an anti-commutation relation for the fields, thus the fermionic character of the quantum field is confirmed, along with all properties that characterize a fermionic field, such as Pauli's exclusion principle and obey Fermi-Dirac's statistics \cite{AHL2,AHL4}.}

{In order to use the Elko field as the matter content in the universe, we will work with its classical formulation, or in terms of average values of its quantum field in a classical background.}
Also, in a curved homogeneous and isotropic space-time we assume that the Elko field is filling all the space homogeneously \cite{BOE1,BOE2,BOE5,FABBRI,FABBRI1,BOE3,BOE4,BOE6,GREDAT,
BASAK,BOE7,WEI,basak,sadja,kouwn,sajf,saj,js,sa,asf}, thus we can assume that it can be split into a time dependent part and a flat space-time dependent part, which carries all the spinor structure, namely, $\lambda(x^\mu)=\phi(t) \xi({\bf x})$, such that $\xi$ stands here, for simplicity, for one of the four kinds of Elko discussed above, normalized as $\stackrel{\neg}{\xi}\xi =\pm 1$.  {A convenient choice for the bare spinor $\xi$ and its dual $\xib$ is:
\begin{eqnarray}
\xi={1\over \sqrt{2}}\left(\begin{array}{c}
\pm i  \\ 
0 \\ 
0 \\ 
1
\end{array} \right)\hspace{0.8cm}
\xib={1\over \sqrt{2}}\Big(i,\;0\; 0,\; \mp 1 \Big)
\end{eqnarray}
We choose to work with a positive norm spinor. Also, since the spinor $\xi$ is constant, we will refer to Elko field just as $\Phi(t)\equiv\phi(t)\xi$.}

Another important characteristic that Elko fields carries is that due to its mass dimension one the possible self-couplings of the field are limited. In particular, for Elko fields, the only two allowed self-couplings are of the type ${1\over 2}m^2\stackrel{\neg}{\lambda}\lambda$ and ${1\over 4}\alpha(\stackrel{\neg}{\lambda}\lambda)^2$. There are also the possibility to couple it to a Higgs field \cite{AHL1,AHL2,AHL3}.

In this paper we study the Elko field coupled to gravity in a Einstein-Cartan framework following recent results \cite{kouwn,sajf}, and we show how it can be the responsible for all phases of the universe. Appendix A contain a brief derivation of the main equations. The action for the Elko field coupled to gravity in a homogeneous and isotropic metric has been already presented in the literature \cite{FABBRI,FABBRI1,BOE6,kouwn,sajf,saj,js,sa,asf}, for both torsion free and torsion coupled equations. The Friedmann equations are given by:
\begin{eqnarray}
H^2&=&\frac{\kappa^2}{3}\bigg[{\dot{\Phi}^2\over 2}+V(\Phi)+{3\over 8}{H^2\Phi^2}+{3\over 4}{H h \Phi^2}\bigg]\nonumber\\
&&+\bigg(1+{\kappa^2\Phi^2\over 8}\bigg)h^2 \,,\label{H2A}
\end{eqnarray}
\begin{eqnarray}
\dot{H}&=&-\frac{\kappa^2}{2}\bigg[{\dot{\Phi}^2}+{3\over 4}{H h \Phi^2}-{1\over 4}{d\over dt}[(H+h)\Phi^2]\bigg] \nonumber\\
&&-3\bigg(1+{\kappa^2\Phi^2\over 8}\bigg)h^2\,, \label{HdotA}
\end{eqnarray}
and the motion equation for the scalar part of the Elko field can be obtained by deriving the first equation and using the second one:
\begin{eqnarray}
\ddot{\Phi}+3H\dot{\Phi}+{V'(\Phi)}-{3\over 4}(H+h)^2\Phi=0\,,\label{eqphiA}
\end{eqnarray}
where
\begin{equation}
h(t)=-\frac{1}{8}{\kappa^2\Phi^2\over(1+\kappa^2\Phi^2/8)}H\;,\label{hfA}\\
\end{equation}
is the only non-null torsion function in the specific case of a homogeneous and isotropic metric \cite{tsam}. After substituting (\ref{hfA}) into Eqs. (\ref{H2A})-(\ref{eqphiA}) and rearranging we are left with:
\begin{equation}
H^2={\kappa^2\over 3}\bigg(1+{\kappa^2\Phi^2\over 8} \bigg)\bigg[{\dot{\Phi}^2\over 2}+V(\Phi)\bigg]\,,\label{H2}
\end{equation}
\begin{equation}
\dot{H}=-{\kappa^2\over 2}\bigg(1+{\kappa^2\Phi^2\over 8} \bigg)\bigg[\dot{\Phi}^2-{1\over 2}{H\Phi\dot{\Phi} \over (1+\kappa^2\Phi^2/8)^2} \bigg]\,,\label{Hdot}
\end{equation}
\begin{equation}
\ddot{\Phi}+3H\dot{\Phi}+{dV(\Phi)\over d\Phi}-{3\over 4}{H^2\Phi\over (1+\kappa^2\Phi^2/8)^2}=0\,,\label{phiElko}
\end{equation}
together the equations for energy density and pressure \cite{sajf}:
\begin{equation}
\rho={\dot{\Phi}^2\over 2}+V(\Phi)+{3\over 8}{H^2\Phi^2\over (1+\kappa^2\Phi^2/8)}\,,\label{rhoElko}
\end{equation}
\begin{eqnarray}
p&=&{\dot{\Phi}^2\over 2}-V(\Phi)-{3\over 8}{H^2\Phi^2\over (1+\kappa^2\Phi^2/8)}- {1\over 4 }{\dot{H}\Phi^2\over (1+\kappa^2\Phi^2/8)}\nonumber\\ &&-{1\over 2}{H\Phi\dot{\Phi}\over (1+\kappa^2\Phi^2/8)^2}\,.\label{pphiElko}
\end{eqnarray}

Notice that the structure of such system of equations is much richer than those corresponding to a standard scalar field, Eqs. (\ref{eqH})-(\ref{pphi}). For this reason the Elko field can be a good candidate to drive not only the inflationary phase of the universe, but also the subsequent phases, as dark matter evolution and accelerated expansion. In which follows we will considerer the above set of equations for each of these eras.

\section{Numerical results}

{In this section it is presented the numerical results concerning tree different phases of expansion of the universe, namely the inflation, dark matter evolution and late time acceleration. A few more details on the numerical equations and values for parameters used are given in Appendix B.}

\subsection{Chaotic Elko inflation}

\begin{figure}[t]
\centering
\includegraphics[width=0.47\textwidth]{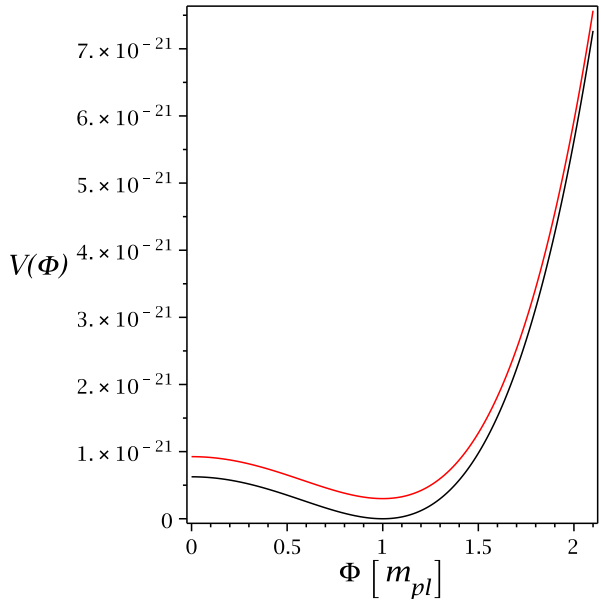}
\caption{Potential (\ref{potential}) (in units of $m_{pl}^4$) constructed with $v_0=0$ (black line) or shifted by $v_0=3.0\times 10^{-22}$ (red line). }
\end{figure}

Now let us start considering the Elko field as a candidate to inflaton field in the universe. We will consider a symmetry breaking potential type plus a constant $v_0$, namely:
\begin{equation}
V(\Phi)=v_0+\Lambda^4\Bigg(1-\frac{\Phi^2}{\sigma^2}\Bigg)^2 = V_0 - \frac{1}{2}\mu^2\Phi^2 + {\alpha\over 4}\Phi^4\,,\label{potential}
\end{equation}
where $V_0 = v_0+\Lambda^4$, $\mu=2\Lambda^2/\sigma$ and $\alpha=4\Lambda^4/\sigma^4$, with $\Lambda$, $\sigma$ and $v_0\ll \Lambda^4$ positive constants. It is well known that such kind of potential represents a particle of physical mass $m=\sqrt{2}\mu$ and has a minimum at $\Phi=\sigma$. Such minimum is zero if $v_0=0$, as showed in Figure 1 (black line), or shifted by $v_0$, as showed in Figure 1 (red line).

Following the chaotic inflationary model by Linde \cite{lindePLB} we consider the pre-inflationary phase of the universe composed by Elko fields distributed chaotically over all the space. In particular we consider that most of fields satisfies\footnote{Notice that a classical description of the evolution of the universe is possible for an energy density satisfying $\rho \ll m_{pl}^4\sim 10^{76}$GeV, thus our only requirement is $V(\Phi)\ll m_{pl}^4$ which may be achieved if $\alpha \ll 1$ for the potential (\ref{potential}). From now on we will write explicitly $\kappa^2=8\pi G=8\pi/m_{pl}^2$.} $\Phi \gtrsim m_{pl}\approx  10^{19}$GeV {and its time variation is greater than the variation of $\Phi$ in a Hubble time, $\dot{\Phi}\gg H\Phi$, which means it is important during inflation}. If the inflation occurs at about $t\sim H^{-1}\approx 10^{-34}$s $\approx 10^{-10}$GeV$^{-1}$ we have $H\ll m_{pl}$, thus the last terms of (\ref{Hdot}) and (\ref{phiElko}) can be discard and we have
\begin{equation}
H^2= \frac{8\pi}{3m_{pl}^2}\bigg(1+{\pi\Phi^2\over m_{pl}^2} \bigg)\bigg[\frac{\dot{\Phi}^2}{2}+ V(\Phi)\bigg]\,,\label{eqHe}
\end{equation}
\begin{equation}
\dot{H} \approx-\frac{8\pi}{2m_{pl}^2}\bigg(1+{\pi\Phi^2\over m_{pl}^2} \bigg)\dot{\Phi}^2\,,\label{eqHdote}
\end{equation}
\begin{equation}
\ddot{\Phi}+3H\dot{\Phi}+V'(\Phi)\approx 0\,,\label{eqphie}
\end{equation}

\begin{figure*}[t]
\centering
\includegraphics[width=0.47\textwidth]{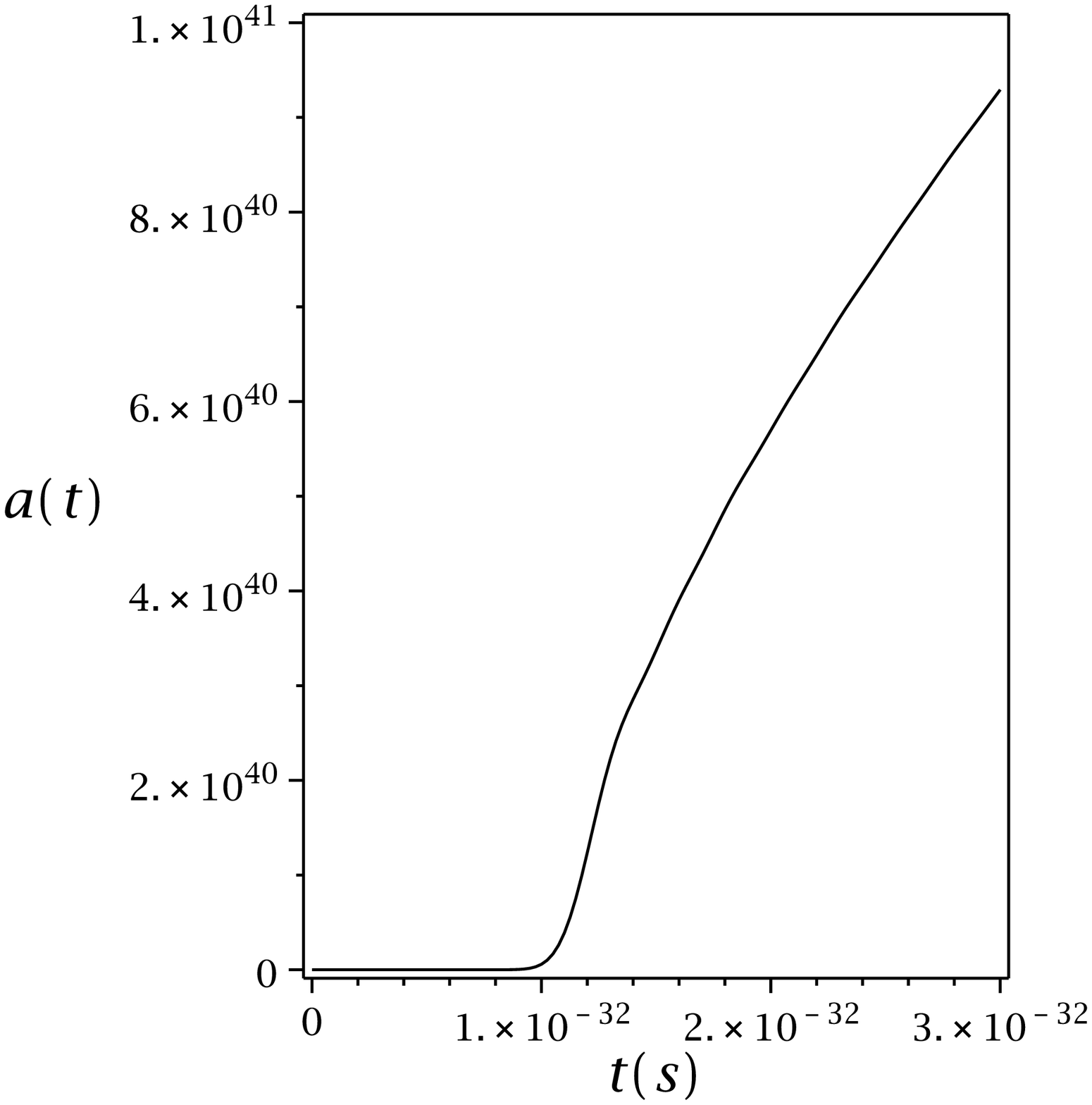}
\hspace{0.3cm}
\includegraphics[width=0.47\textwidth]{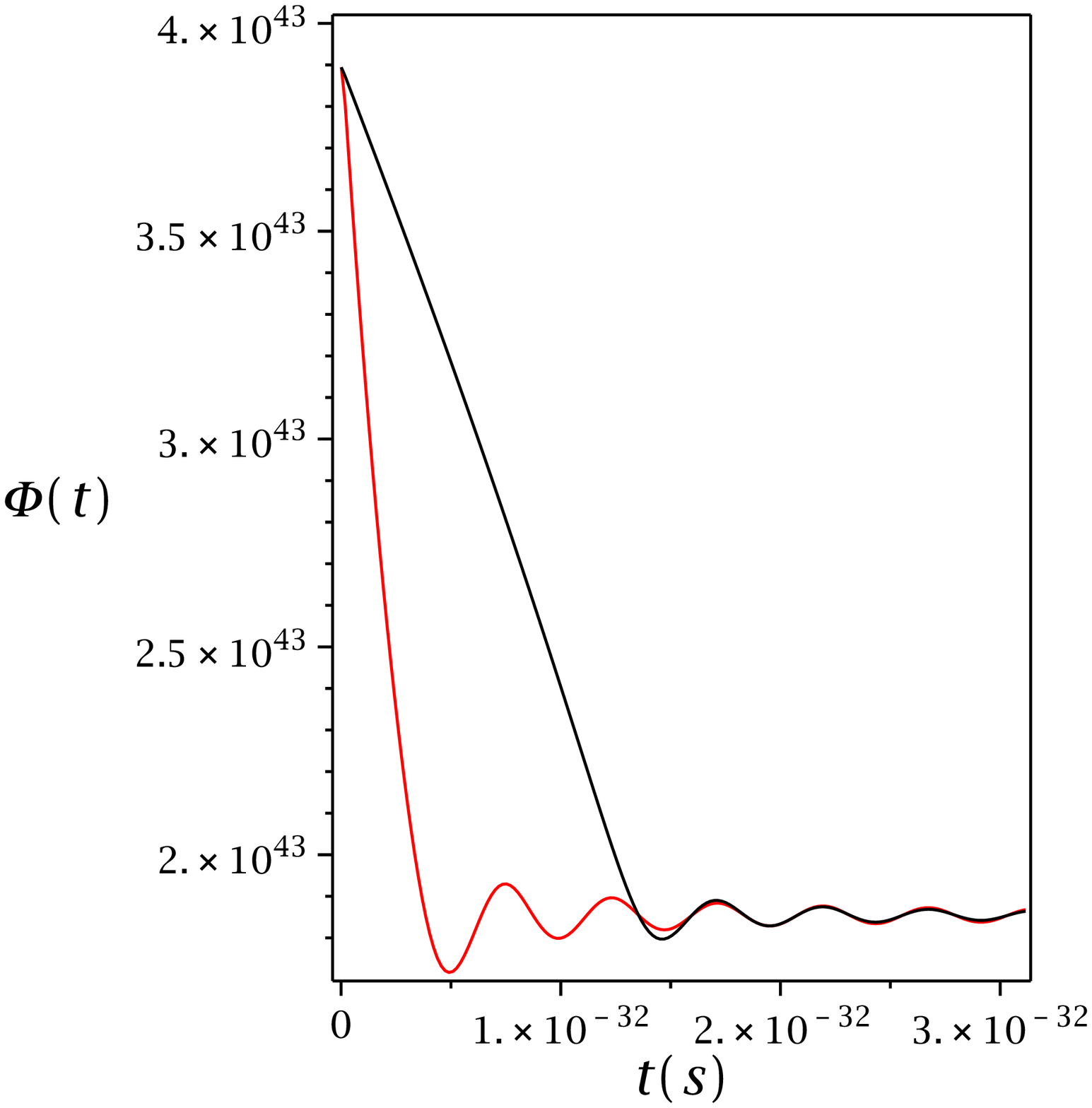}\\
\hspace{2.1cm}(a)\hspace{8.6cm}(b)
\includegraphics[width=0.47\textwidth]{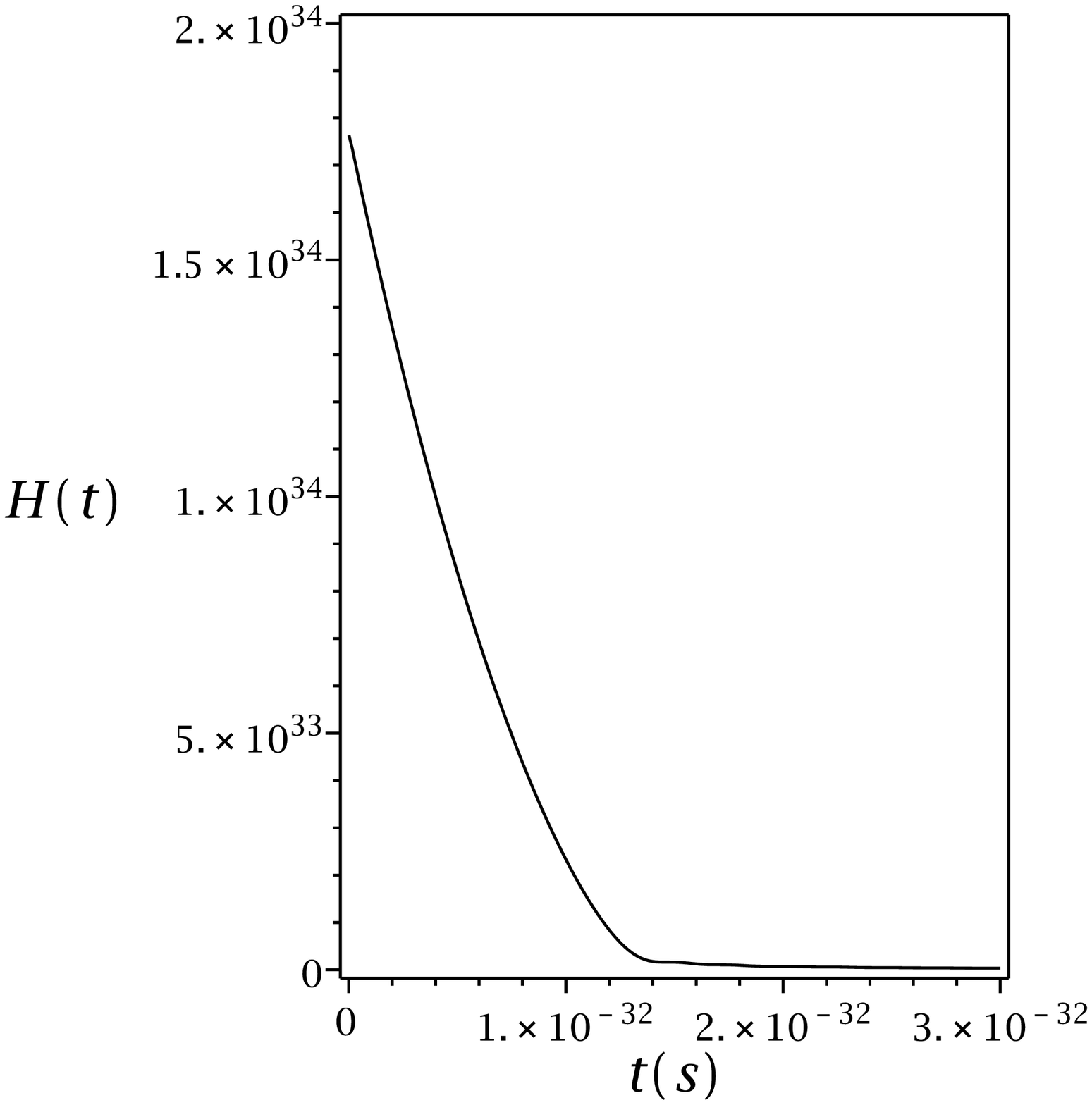}
\hspace{0.3cm}
\includegraphics[width=0.47\textwidth]{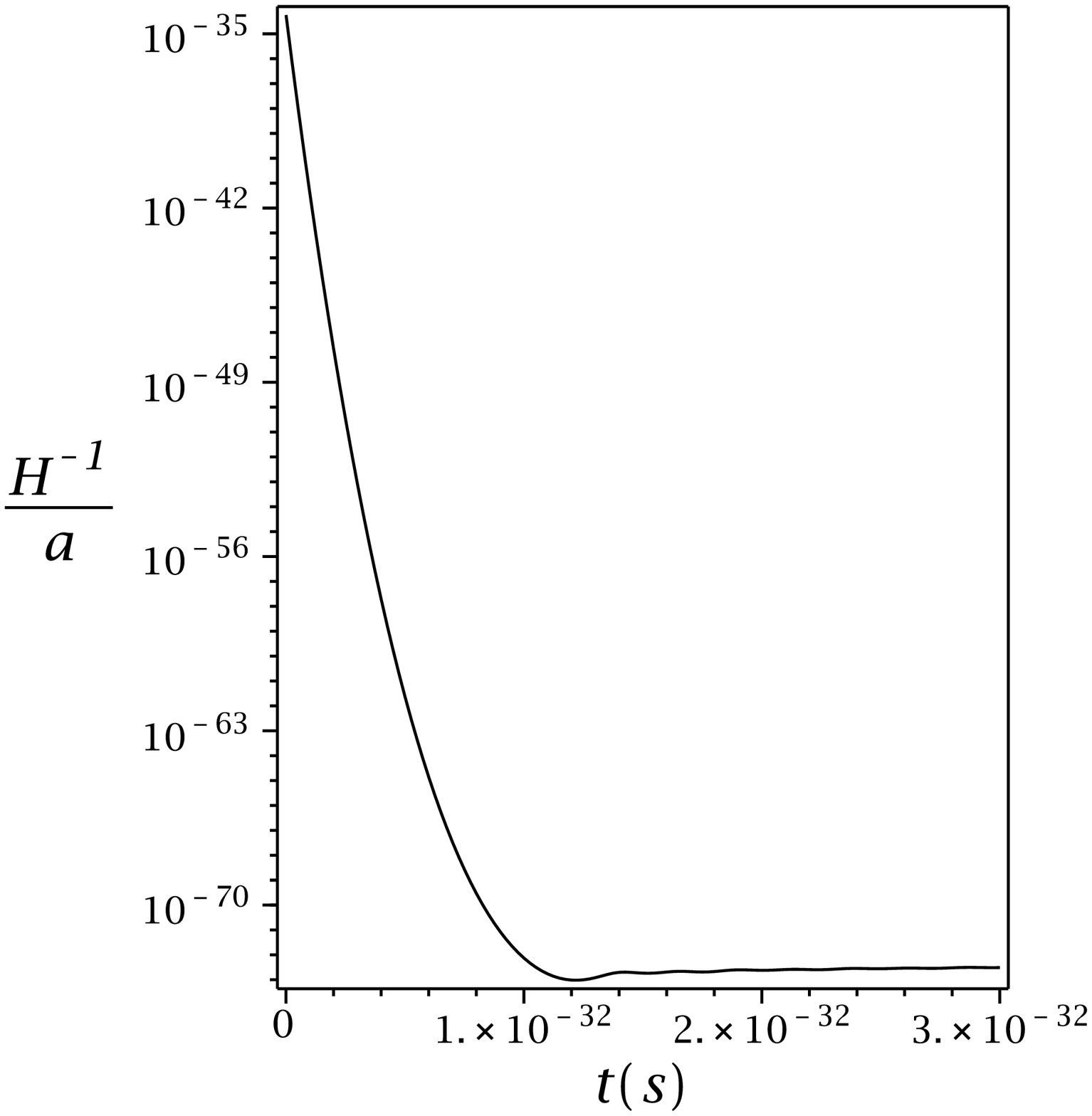}\\
\hspace{2.1cm}(c)\hspace{8.6cm}(d)
\caption{ Numerical results for the parameters $a(t),\, \Phi(t),\,H(t)$ and $H(t)^{-1}/a(t)$ during the inflationary phase, from $t_i=1.0\times 10^{-35}$s up to $t_f=3.0\times 10^{-32}$s, obtained with the parameters $\Phi_i=2.1m_{pl}$, $\sigma = 1.0m_{pl}$, $\Lambda=5\times 10^{-6}m_{pl}\simeq 6.1\times 10^{13}$GeV and $v_0=0$. (a) - Evolution of the scale factor $a(t)$ with initial condition $a_i=1$. (b) - Decay of the field $\Phi(t)$ (in units of s$^{-1}$) for initial values above and also $d\Phi(0)/dt\equiv \dot{\Phi}_i=0$ for the equation (\ref{phiElko}) in the presence of the last term (black line) and in the absence of the last term (red line). (c) Evolution of $H(t)$ (in units of s$^{-1}$). (d) Evolution of $H(t)^{-1}/a(t)$. }
\end{figure*}

\begin{figure*}[t]
\centering
\includegraphics[width=0.47\textwidth]{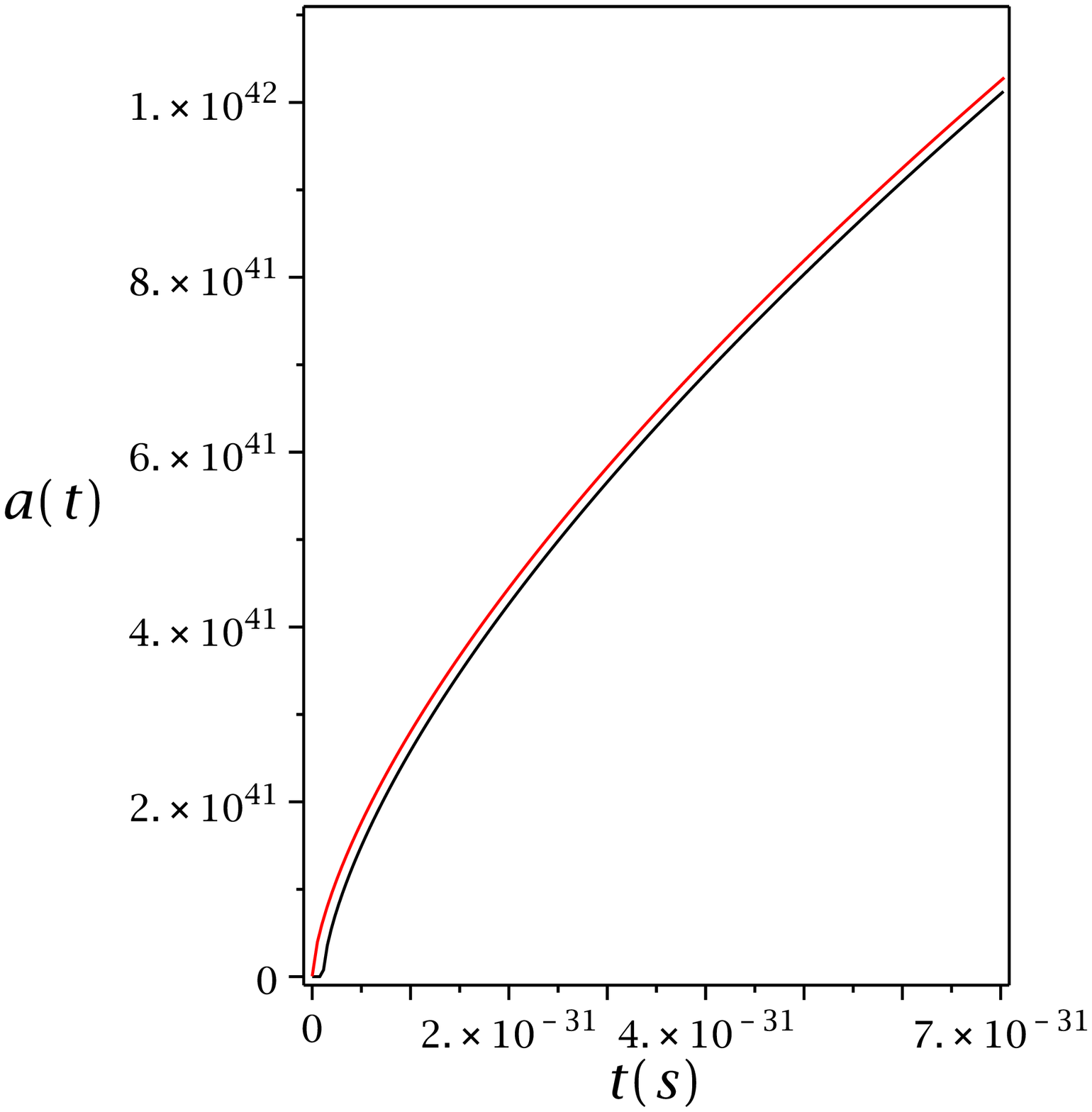}
\hspace{0.3cm}
\includegraphics[width=0.47\textwidth]{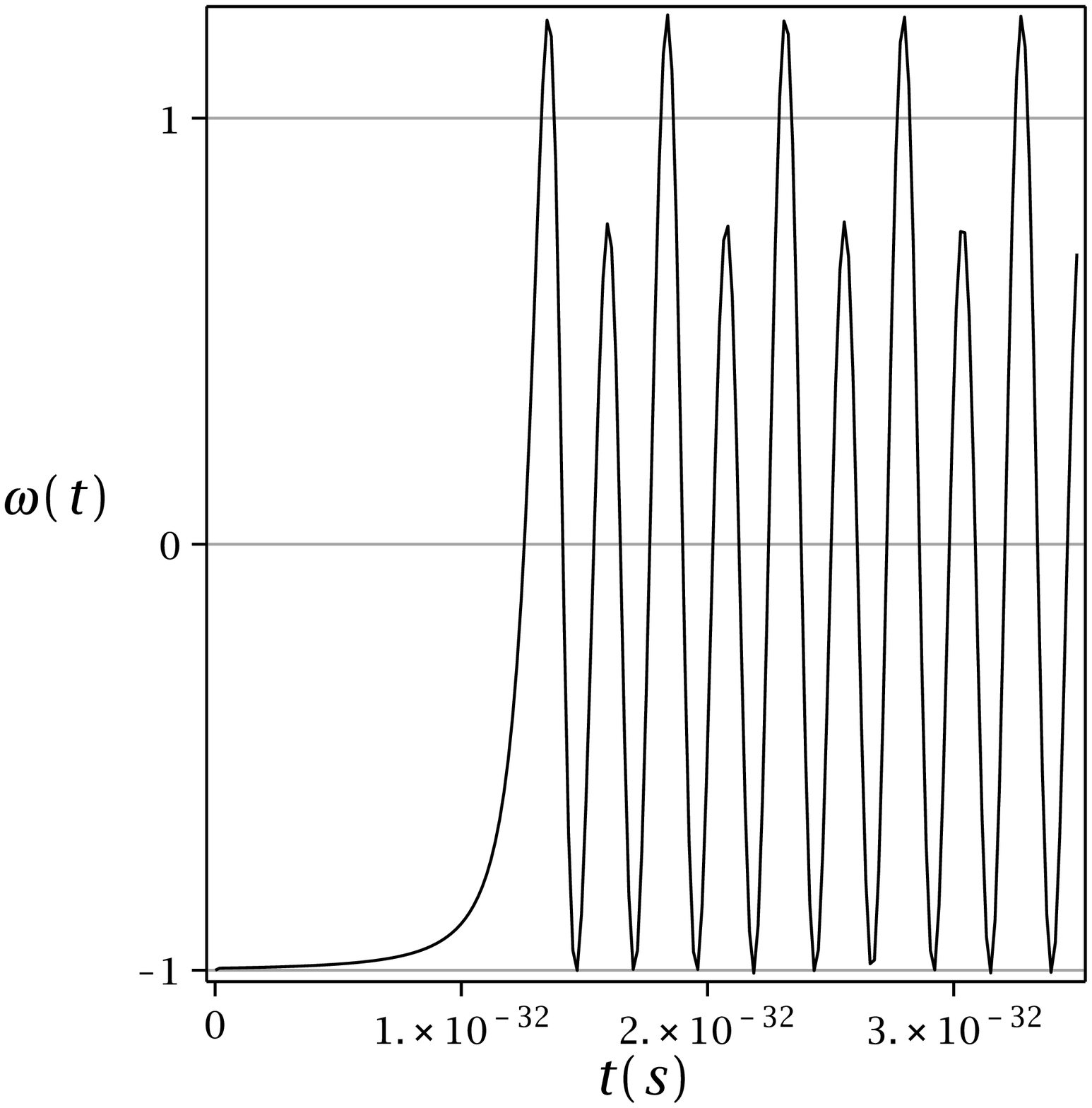}\\
\hspace{2.1cm}(a)\hspace{8.6cm}(b)
\includegraphics[width=0.47\textwidth]{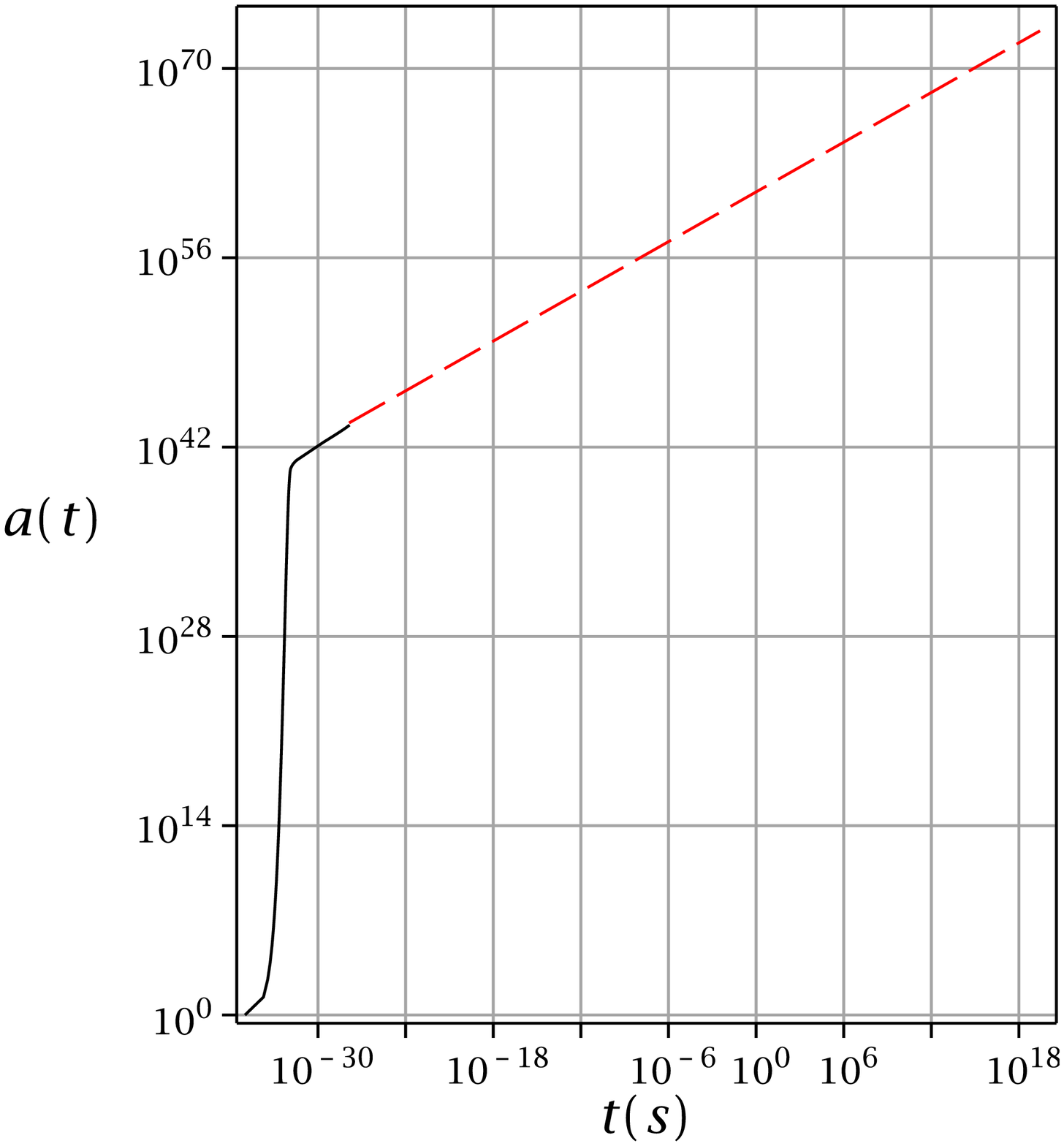}
\hspace{0.3cm}
\includegraphics[width=0.47\textwidth]{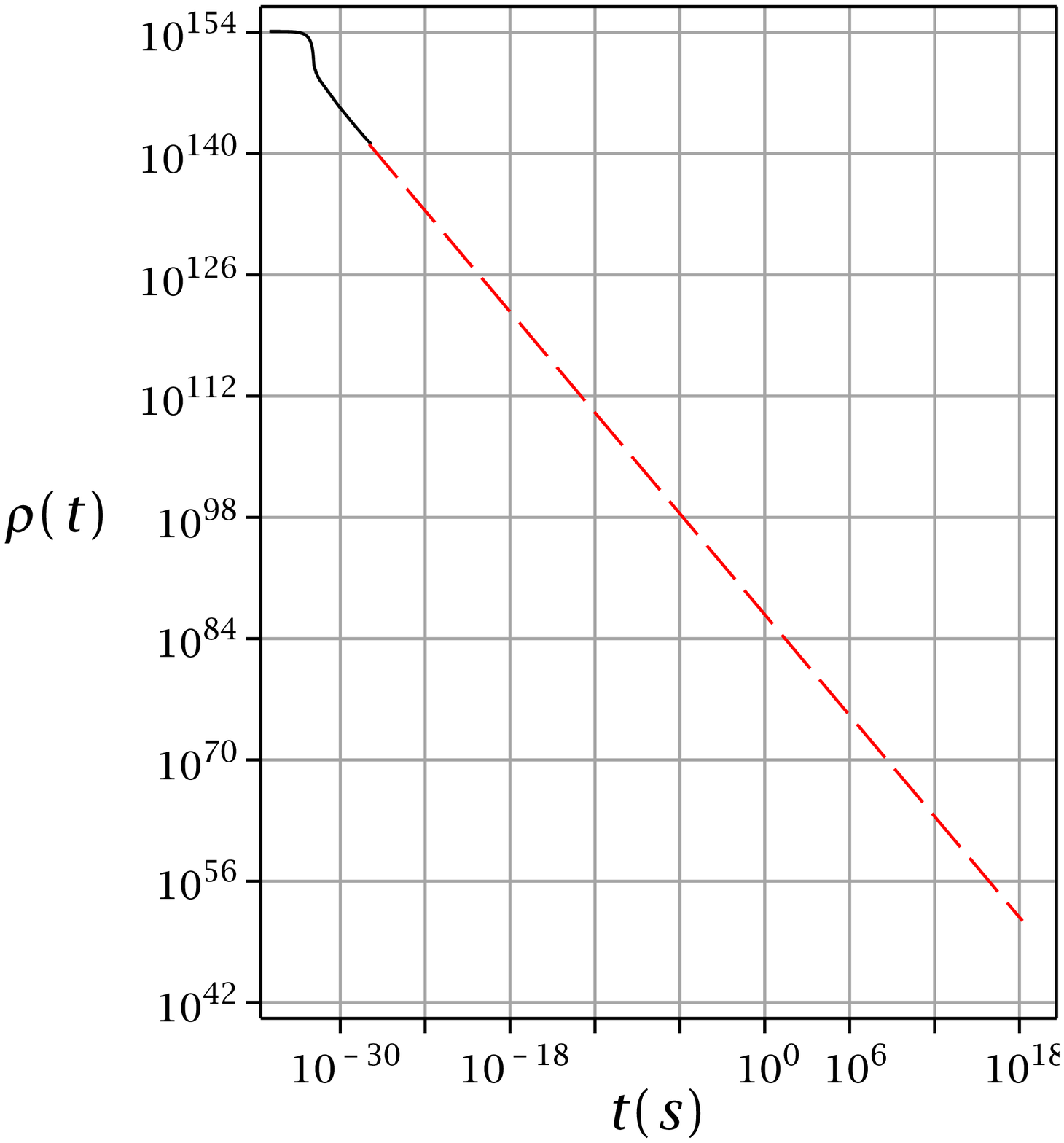}\\
\hspace{2.1cm}(c)\hspace{8.6cm}(d)
\caption{ (a) - Numerical result for the scale factor evolution for a long time (black line). In red line we plot the function $a(t)=a_0t^{2/3}$ for comparison, with $a_0=1.30\times 10^{62}$. (b) - Numerical analysis for $\omega(t)$ from (\ref{eqstado}). The average oscillation is around $\omega=0$. (c) - Extrapolation of the curve $a(t)=a_0t^{2/3}$ up to present time (red dashed line) and the numerical result for $a(t)$ (black line) in logarithm scale. (d) - Numerical result for the evolution of the energy density (\ref{rhoElko}) during the inflation (black line) and extrapolated function up to present time (red dashed line) (in units of s$^{-4}$).}
\end{figure*}

Except for the term inside curl brackets (that is important only in the limit $\Phi  \gtrsim m_{pl}$) these expressions are quite equivalent to (\ref{eqH})-(\ref{eqphi}). Such impressive result shows that all successful inflationary models with a (unknown) scalar field can now be used here, where the inflaton field is the Elko field, a much more physically reasonable field that is a natural candidate to dark matter particle. In particular, the slow-roll condition (\ref{eqVl}) responsible for the beginning of the inflation is maintained (considering $V\gg \dot{\Phi}^2$), since that the ratio $\dot{H}/H^2$ will exactly cancel the term inside curl brackets, which ensures that such inflationary model starts exactly as in the standard scalar field model. For the second slow-roll condition (\ref{eqVll}) it can be shown that it is written as:
\begin{equation}
\vert\eta(\Phi)\vert  \simeq \Bigg\vert\frac{1}{\kappa^2}\frac{V''(\Phi)}{\big(1+\kappa^2\Phi^2/8\big)V(\Phi)}\Bigg\vert \ll 1\,,\label{eqVllElko}
\end{equation}
so that when $\Phi\ll m_{pl}$ the expression (\ref{eqVll}) is recovered and when $\Phi\gg m_{pl}$ we have $\eta \simeq (m_{pl}^4/8\pi^2\Phi^2)(V''/V)$.

After inflation start the field must decay according to the general equations (\ref{H2})-(\ref{phiElko}). 

We will present some numerical results for the complete system of equations (\ref{H2})-(\ref{pphiElko}) taking the particular potential (\ref{potential}) in order to illustrate the validity of the model (see Appendix B for some details). Before that, let us make some estimates on the parameters of $V(\Phi)$ in order to reproduce viable models of inflation. First notice that the parameter $\sigma$ is an energy scale that characterize the final evolution of the field $\Phi$, since that the initial field $\Phi_i > \sigma$ rolls down to the bottom of the potential and when $\Phi \to \sigma$ the potential vanish, $V(\Phi\to \sigma)=0$ and the inflationary mechanism ends. In this sense it will characterize the number of e-foldings of the inflation. It is also directly related to the physical mass of the field, together the parameter $\Lambda$, from (\ref{potential}). The parameter $\Lambda$ characterizes the total potential energy of the field and a strong constraint into it is $V(\Phi_i)\ll m_{pl}^4$, which guarantees that the field energy is bellow Planck scale. We also expect inflation to occur after a initial time $t_i \sim 10^{-35}$s, once we do not know exactly what happens before such time, and finish at about $t_f\sim 10^{-32}$s. After that the universe expands according to the standard model of cosmology. 

Now we will present the numerical results for the following set of parameters: $v_0=0$, $\Phi_i=2.1m_{pl}$, $\sigma = 1.0m_{pl}$ and $\Lambda=5\times 10^{-6}m_{pl}\simeq 6.1\times 10^{13}$GeV. For our model we have at the beginning of inflation:
\begin{equation}
H^2(t_i) \simeq \frac{8\pi}{3m_{pl}^2}\bigg(1+\frac{\pi\Phi_i^2}{m_{pl}^2}\bigg)V(\Phi_i)\simeq 9.04\times 10^{-19}m_{pl}^2\,,
\end{equation}
which leads to $t_i=H^{-1}\simeq 5.7\times 10^{-35}$s, a very reasonable value. Notice also that for such $\Phi_i$ we have $V(\Phi_i)\simeq 7.2\times 10^{-21}m_{pl}^4$, which guarantees the condition $V(\Phi_i)\ll m_{pl}^4$. Also, for such parameters we have $\epsilon\approx 0.24$ and $\eta\approx 0.011$ for the slow roll parameters (\ref{eqVl}) and (\ref{eqVllElko}). The physical mass for such field is $m=\sqrt{2}\mu = 7.1\times 10^{-11}m_{pl}\simeq 8.6\times 10^{8}$GeV and the value of the self-coupling constant is $\alpha\simeq 2.5\times 10^{-21}$.

\begin{figure*}[ht]
\centering
\includegraphics[width=0.97\textwidth]{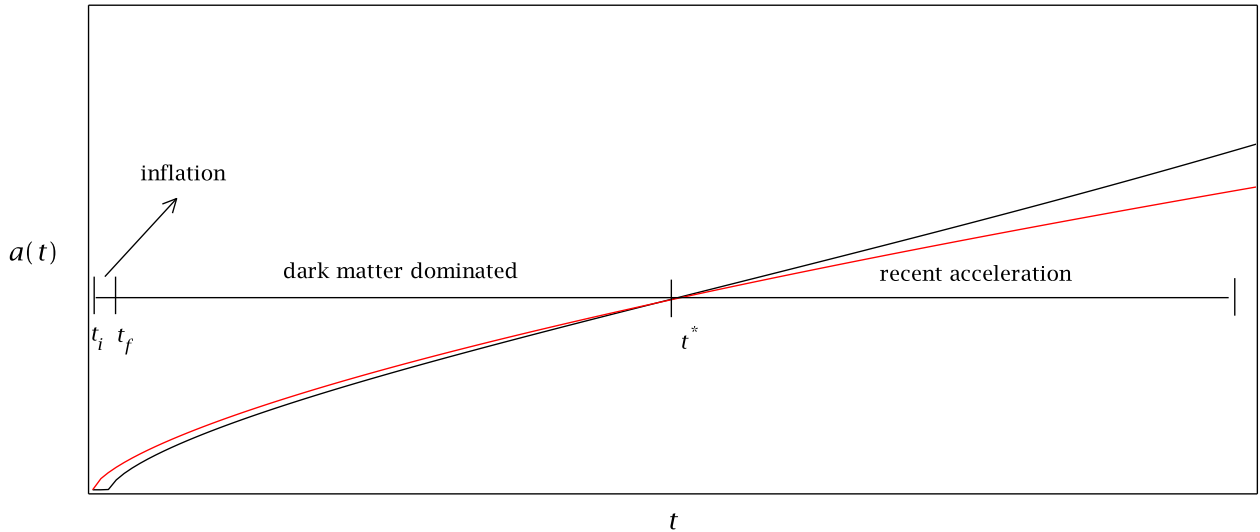}
\caption{ (a) - Evolution of the scale factor with a potential containing a shift term $v_0$ (black line) and the evolution proportional to $t^{2/3}$ (red line), for comparison. The crossing of the curves occurs at $t^*$ that goes to infinity when $v_0\to 0$.}
\end{figure*}

Figure 2 (a) shows the numerical result for the scale factor evolution for the parameters indicated in the figure. We see that the scale factor grows for several order of magnitude from an initial time $t_i\sim 0$ up to a final time of about $t_f \sim \simeq 1.25\times 10^{-32}$s as expected ($t_f$ characterize the end of inflation, and we are using $t_i = 10^{-35}$s in the numerical analysis). The scale factor grows from $a_i=1$ to $a_f\simeq 2\times 10^{40}$ during the inflationary phase, which leads to an e-folding number of about $N=93$, in reasonable accord to modern theories of inflation. We have also verified numerically that the choice of the initial value of the field $\Phi_i$ alters drastically the amount of e-foldings without to alter significantly its duration, which is more sensitive to changes on the $\Lambda$ parameter. Greater the values of $\Lambda$ smaller the duration of inflation. After this we observe that the exponential evolution stops and the universe evolves in a non-accelerating phase. 

Figure 2 (b) shows the numerical behaviour for the field $\Phi(t)$. It starts from $\Phi_i=2.1m_{pl}\simeq 3.9\times 10^{43}$s$^{-1}$ and $\dot{\Phi}_i=0$ and rolls down to the bottom of the potential, oscillating around its minimal value of $\bar{\Phi}=\sigma = 1.0m_{pl}$ in this case. The black line shows the numerical result for the complete Eq. (\ref{phiElko}), including its last term, and red line shows the behaviour in absent of last term. It is clear that the presence of the last term causes a kind of damping while the field rolls down, delaying its fall to the bottom of the potential. Such behaviour is an analogous to that one of the phenomenological term proportional to $\Gamma$ for the standard scalar field (\ref{eqphiG}).

Figure 2 (c) shows the evolution of $H(t)$ decreasing abruptly after inflation and Figure 2 (d) the evolution of $H(t)^{-1}/a(t)$, showing that its time derivative are negative up to about $t_f \sim 1.20\times 10^{-32}$s, as required by (\ref{condinflation}).

\subsection{Dark matter evolution}

Now let us analyse how is the behaviour of the Elko field after inflation. As already indicated by the Figure 1 (a), after inflation the scale factor evolution change its concavity and another phase of evolution takes place. It is well known from standard model of cosmology that the next phase is a radiation dominated universe, scaling as $t^{1/2}$, since the hot universe just after inflation is the responsible for the CMB radiation observed today, as the universe cools. After radiation the universe enters a phase dominated by dark matter, scaling as $t^{2/3}$, the pressure of matter is null, indicating that particles stops its collisions and galaxies and cluster of galaxies start to form.

Contrary to the standard inflationary model constructed with a scalar field, where the inflaton field must decay to zero during the reheating in order to radiation and dark matter start to dominate, in our model constructed with the Elko field the evolution continues driven by equations (\ref{H2})-(\ref{pphiElko}). In Figure 3 (a) we show the numerical analysis for the evolution of the scale factor for a very long time after inflation (black line), with the same parameters before. We also plotted in the same figure a scale factor evolution of the form $t^{2/3}$ (red line) for a matter dominated universe, just for comparison. It is clear from the analysis of the figure that the evolution after inflation is exactly like a matter dominated universe, as expected, since the Elko field has exactly the physical characteristics of a dark matter particle. 

Another way to analyse such evolution after inflation is by mean of the equation of state parameter:
\begin{equation}
\omega(t)=\frac{p(t)}{\rho(t)}\,.\label{eqstado}
\end{equation}  
A numerical analysis for $\omega(t)$ taking (\ref{rhoElko})-(\ref{pphiElko}) is showed in Figure 3 (b). After start from $\omega=-1$, a vacuum type equation of state parameter, responsible for the acceleration during inflation, the equation of state parameter starts to oscillate at the end of inflation. The first amplitude grows from $-1$ to slightly above $1$ while the second grows from $-1$ to above 0. It is clear that the average oscillation is around 0, showing that during all the future evolution the equation of state parameter is of dust type, or pressureless, as required by a dark matter field.

Due to numerical limitations (long time of calculations), we cannot cover all the time scale up to present time for the evolution of the scale factor. Nevertheless, having the function that plots the evolution after inflation, namely $a(t)=a_0t^{2/3}$, we can extrapolate such function up to present time. This is shown in Figure 3 (c) (red dashed line) in a logarithm scale. {In black line we shown the numerical result for $a(t)$ up to $t\sim 10^{-28}$s. Notice that the end of time scale is about $\simeq 10^{18}$s, exactly today. The scale factor growth to about $10^{74}$, exactly as predicted by standard model of cosmology \cite{kolb,weinberg,uzan}.}

Another very important quantity concerning the Elko field during its evolution is its energy density $\rho$, given by (\ref{rhoElko}). During the inflationary phase we expect it to be nearly constant. At the end of inflation the energy density of the universe must be about $10^{70} - 10^{65}$g/cm$^{3}$, depending exactly when the inflation ends \cite{kolb,weinberg,uzan}. In Figure 3 (d) we plot in a logarithm scale the numerical result for the energy density (\ref{rhoElko}) during inflation and just after it. {It is clear that the energy density is nearly constant at beginning  and then  decreases abruptly after inflation ends. If the inflation ends at about $t_f\sim 10^{-32}$s, the corresponding energy density at this time is about $10^{150}$s$^{-4} \sim 10^{53}$GeV$^4\simeq 10^{70}$g/cm$^3$, in a good agreement to estimates of standard model. This is an important constraint of the model. Also, if we extrapolate the curve of the energy density up to present time, showed in red dashed line of Figure 3 (d), we obtain an energy density for the present time of about $10^{52}$s$^{-4} \sim 10^{-45}$GeV$^4\simeq 10^{-28}$g/cm$^3$. The estimated value for the present time is about $10^{-27}$g/cm$^3$, also indicating a good agreement to standard model, since we must also add baryonic matter and radiation to current model of cosmology.}

\subsection{Recent cosmic acceleration}  

As a final remark let us see how we can address to this model the recent accelerated evolution of the universe. Looking for the potential from Figure 1 with $v_0=0$ (black line), it is clear that as the field rolls down to its average value $\bar{\Phi}=\sigma$ and goes to the bottom of the potential we have $V(\bar{\Phi})\to 0$ and the evolution follows as $t^{2/3}$. But we have noticed that, if the potential is slightly shifted by a small $v_0$, we have $V(\bar{\Phi})\to v_0$ and such term will act exactly as a cosmological constant term at the future evolution of the universe. 

Figure 4 shows the numerical analysis in a arbitrary time scale for the behaviour of scale factor in the presence of such term.  It is clear that the universe starts in an inflationary phase at $t_i$ that ends at $t_f$, then it passes to a phase nearly proportional to $t^{2/3}$ from $t_f$ to $t^*$ and then it starts a new accelerating phase after $t=t^*$. In the numerical analysis we have used $v_0=1.0\times 10^{-28}m_{pl}^4$, but the value of $t^*$ is not realistic here, since the acceleration starts only in the future. Even in this qualitative analysis, it is also obvious that the actual accelerating phase is much more smooth than the inflationary evolution. We have also verified numerically that the instant of time where the acceleration start to dominate is dependent just on $v_0$. When $v_0\to 0$ the crossing between the two curves of Figure 4 will occur each more in the future, namely $t^*\to \infty$. The precise value of $v_0$ must be constrained by observations.

\section{Concluding remarks}
In this paper we have studied a cosmological scenery where the mass dimension one Elko field subject to a symmetry breaking potential is the only matter content of the universe. Following the chaotic inflationary model, for an Elko field characterized by a time evolution represented by $\Phi(t)$ at an initial energy scale $\Phi_i > m_{pl}$, we have obtained numerically that the dynamical evolution of the Elko field rolling down to the bottom of the symmetry breaking potential has the desired properties of an inflaton field. The slow-roll conditions for the system were obtained in the limit $H\ll m_{pl}$ and it was showed that they are satisfied for the initial conditions in the present case.

{Several important aspects can be found in the model after numerical results. First, a nearly exponential growth of the scale factor from $t_i\sim 10^{-35}$s up to $10^{-32}$s leads to an inflation of about $N=93$ e-foldings. After inflation, the field enters a dark matter era evolving as $t^{2/3}$, oscillating around the minimum of the potential, resting at an energy of about $m_{pl}$, where the potential energy is null. The energy density at the end of inflation is the expected one according to standard model, about $10^{70}$g/cm$^3$. By making an extrapolation of the curve $a(t)=a_0t^{2/3}$ up today, we have found that the scale factor growth to about $10^{74}$ order of magnitude, also in good agreement to standard model. Additionally, the energy density of the field for present time is $\rho\sim 10^{-28}$g/cm$^3$, just one order of magnitude bellow the predicted by standard model.  Finally, if the potential is slightly shifted by a constant term, the scale factor enters a new phase similar to a cosmological constant dominated universe, reproducing the present accelerated phase of the universe. Such constraints on different epochs are important results of the model, which must be better constrained with observations and also including radiation and baryonic matter to the model.}

Another interesting properties naturally follows from the system of equations. First, the very similar form of the equations that governs the time evolution of Elko field with the standard scalar field in the limit $H\ll m_{pl}$. Also, the presence of a correction term similar to a kind of damping term {present} in the standard scalar field model of inflation. Here such term appears naturally. While the field is rolling down to the bottom of the potential its equation of state parameter goes from $-1$ to an average oscillation around zero, which guarantees a dark matter evolution after the initial exponential growth.

We have also verified numerically that greater the value of the initial field $\Phi_i$ greater the number of e-foldings of the inflationary expansion. The $\sigma$ parameter also alters the number of e-foldings. Greater the value of the parameter $\Lambda$ into the potential lesser the duration of the inflationary exponential growth.

{As a final remark concerning the inflation driven by Elko, if we consider the Elko field just as a classical field, the numerical solutions shows that it has the desired properties to drive inflation, but the physical mechanism to this can be better understood if we treat the Elko field as a quantum one. The complete Elko field $\lambda=\Phi(t)\xi$ can be treated as an average value $\langle \lambda \rangle$ of the quantum field, and the above set of equations must remain valid. 

{A possible way to understand the inflationary phase as a result of the Pauli exclusion principle is as follows. In pre-inflationary phase the particles are filling the energy states according to Pauli exclusion principle, just one particle in the ground state, one particle in the first excited state and so on, with a large energy spacing among them. But all particles are rolling down to the bottom of potential, each one trying to occupy the minimal energy state, while the degeneracy pressure prohibits particles of greater energy from occupying lower energy states. Such system can stay in equilibrium in this configuration, as occurs in a neutron star, or all particles can  nearly reach the ground state energy (with a small energy spacing among them) provided that the volume of the system increase significantly, as occurs after inflation. It is well known that the energy spacing in some quantum systems are inversely proportional to its volume. The effect of the degeneracy pressure is to enlarge the system, given rise to inflation, where the volume of the whole system increases in order to allow the particles to occupy nearly the lowest energy state. Thus the Pauli exclusion principle act as a repulsive force and makes the whole system to expand, given rise to inflation.} Notice that such quantum interpretation is not possible for a bosonic scalar field, the main ingredient in the standard model of inflation.

This and other properties deserves future investigations in order to place the Elko field as a good candidate to {drive inflation and other phases of evolution of the universe}.

\onecolumngrid

\section*{APPENDIX}

\subsection{Elko in Einstein-Cartan framework}

The action for Elko field in a general Einstein-Cartan framework is:
\begin{equation}
S = \int d^4 x \sqrt{-g} \left[ -\frac{1}{2\kappa^2}\tilde{R}+{1\over 2}g^{\mu\nu}\tilde{\nabla}_\mu \lambdab\tilde{\nabla}_\nu \lambda -V(\lambdab \lambda) \right] \,.
\label{action}
\end{equation}

The flat FRW metric with a lapse function $N(t)$ can be written in terms of vierbein:
\begin{equation}
g_{\mu\nu}=e_\mu^{\;\;a}e_\nu^{\;\;b}\eta_{a b},
\end{equation}
where $\eta_{a b}=diag(1,-1,-1,-1)$ and $e_\mu^{\;\;a}$ is given by
\begin{equation}
e_\mu^{\;\;a}=[N(t),a(t),a(t),a(t)]\,, \qquad e^\mu_{\;\;a}=[{1\over N(t)},{1\over a(t)},{1\over a(t)},{1\over a(t)}].
\end{equation}
Greek indexes stands for curved spacetime and latin indexes for Lorentz indexes. Dirac matrices $\gamma^\mu$ in curved spacetime are related to $\gamma^a$ in Minkowski spacetime by ${\gamma}^{\mu}={e^{\mu}}_{a}{\gamma}^{a}$, satisfying:
\begin{equation}
\gamma^{\mu}\gamma^{\nu}+\gamma^{\nu}\gamma^{\mu}=2g^{\mu\nu}, \quad \gamma^{a}\gamma^{b}+\gamma^{b}\gamma^{a}=2\eta^{ab}.
\end{equation}

The covariant derivatives of a spinor and its dual are defined as
\begin{equation}
\tilde{\nabla}_\mu \lambda\equiv \partial_\mu\lambda - \tilde{\Gamma}_\mu \lambda\,, \qquad \tilde{\nabla}_\mu \lambdab\equiv \partial_\mu\lambdab + \lambdab\tilde{\Gamma}_\mu\,,
\label{cdd}
\end{equation}
where tilde denotes the presence of torsion. The spin connection ${ \tilde{\Gamma}}_{\mu}$ is given  by \cite{spinor}:
\begin{equation}
\tilde{\Gamma}_\mu=\frac{1}{8}{\omega_{\mu}}^{ab}\left[\gamma_{a},\gamma_{b}\right],
\end{equation}
where ${{\omega_{\mu}}^{ab}} = e^{a}_\nu\partial_\mu{e^\nu}^{b}+e^{a}_\nu\tilde{\Gamma}^{\nu}_{\mu\rho} {e^\rho}^{b}$ and the affine connection containing the contorsion $K_{\;\;\mu\nu}^\rho$ is given by:
\begin{equation}
\tilde{\Gamma}^\rho_{\mu\nu} = \Gamma^\rho_{\mu\nu}+K^\rho_{\;\;\mu\nu}\,,
\end{equation}
where $\Gamma^\rho_{\mu\nu}$ is the standard Christoffel symbol. The contorsion is written in terms of the torsion tensor $T_{\;\;\mu\nu}^\rho$ as:
\begin{equation}
K^\rho_{\;\;\mu\nu}=-{1\over 2}(T^\rho_{\;\;\mu\nu}+T_{\mu\nu}^{\;\;\;\;\rho}+T_{\nu\mu}^{\;\;\;\;\rho} )\,.
\end{equation}
For a homogeneous and isotropic metric in a Riemann-Cartan spacetime, the non-vanishing components of torsion are \cite{tsam}:
\begin{equation}
T_{110}=T_{220}=T_{330} = -T_{101}=-T_{202}=-T_{303}= a(t)^2h(t),
\end{equation}
\begin{equation}
T_{ijk}=2a(t)^3f(t)\varepsilon_{ijk},
\end{equation}
and the functions $h(t)$ and $f(t)$ are general and $\varepsilon_{ijk}$ is the totally antisymetric symbol. The non-vanishing components of the connection are \cite{chee}:
\begin{equation}
\tilde{\Gamma}^0_{00}={\dot{N}\over N}\,,\quad \tilde{\Gamma}^0_{ij}={a\dot{a}+a^2h\over N^2}\delta_{ij}\,,\quad \tilde{\Gamma}^i_{0j}={\dot{a}+ah\over a}\delta_{ij}\,, \quad \tilde{\Gamma}^i_{j0}={\dot{a}\over a}\delta_{ij}\,, \quad \tilde{\Gamma}^i_{jk}=-af \varepsilon_{ijk}\,. 
\end{equation}
The Ricci curvature scalar is:
\begin{equation}
\tilde{R}=-6\Bigg[{1\over aN}{d\over dt}\bigg({\dot{a}+ah\over N}\bigg)+\bigg({ \dot{a}+ah\over aN}\bigg)^2 - {f}^2 \Bigg]\,.
\end{equation}
By assuming the Elko fields as $\lambda=\phi(t) \xi\equiv \Phi(t)$, such that $\xi$ is a constant spinor and $\xib\xi =1$, the lagrangian density reads: 
\begin{equation}
\mathcal{L} = -{1\over N}\bigg({3a\dot{a}^2\over \kappa^2} - {3a^3h^2\over \kappa^2} - {1\over 2}a^3\dot{\Phi}^2-{3\over 8}a(\dot{a}+ah)^2\Phi^2\bigg)-N\bigg({3a^3f^2\over \kappa^2}+{3\over 8}a^3f^2\Phi^2+a^3V(\Phi)\bigg)\,,
\end{equation}
where $V(\Phi)$ is the potential.

Taking the Euler-Lagrange equations with respect to $N(t)$, $a(t)$, $\Phi(t)$, $h(t)$ and $f(t)$ we obtain (setting $N\to 1$ at the end), respectively
\begin{equation}
3H^2={\kappa^2}\bigg[{\dot{\Phi}^2\over 2}+V(\Phi)+{3\over 8}{H^2\Phi^2}+{3\over 4}{H h \Phi^2}\bigg]+3\bigg(1+{1\over 8}\kappa^2\Phi^2\bigg)h^2+3\bigg(1+{1\over 8}\kappa^2\Phi^2\bigg){f^2} \,,\label{H2AA}
\end{equation}
\begin{equation}
-2\dot{H}-3H^2={\kappa^2}\bigg[{\dot{\Phi}^2\over 2}-V(\Phi)-{3\over 8}{H^2\Phi^2}-{1\over 4}{d\over dt}[(H+h)\Phi^2]\bigg] +3\bigg(1+{1\over 8}\kappa^2\Phi^2\bigg)h^2-3\bigg(1+{1\over 8}\kappa^2\Phi^2\bigg){f^2}\,,\label{HdotAA}
\end{equation}
\begin{equation}
\ddot{\Phi}+3H\dot{\Phi}+{dV(\Phi)\over d\Phi}-{3\over 4}\bigg((H+h)^2-{f^2}\bigg)\Phi=0\,,\label{eqphiAA}
\end{equation}
\begin{equation}
h(t)=-{1\over 8}{\kappa^2\Phi^2\over (1+\kappa^2\Phi^2/8)}\bigg({\dot{a}\over a}\bigg)\;, \qquad f(t) = 0\,,\label{hfAA}\\
\end{equation}
where $H=\dot{a}/a$, as usual. Written in this form, the right side of (\ref{H2AA}) and (\ref{HdotAA}) are, respectively, energy density and pressure of the field. Substituting $h(t)$ and $f(t)$ from (\ref{hfAA}) into (\ref{H2AA}), (\ref{HdotAA}) and (\ref{eqphiAA}) we obtain, after some algebraic manipulations, the equations (\ref{H2A}), (\ref{HdotA}) and (\ref{eqphiA}).

\subsection{Numerical analysis of the system of differential equations}

In this Appendix we will briefly present in a few more details the numerical method for solution of the coupled system of differential equations (\ref{H2})-(\ref{pphiElko}) in order to construct the Figures of Section II. We use the DEtools Package from Maple 15 Software, where numerical solutions are found by the method $rkf45-dae$, which is an extension of the $rkf45$ method, which finds a numerical solution using a Fehlberg fourth-fifth order Runge-Kutta method with degree four of interpolation.

For the figures concerning the scale factor $a(t)$ and the field $\Phi(t)$ we use the  equations (\ref{H2}) and (\ref{phiElko}) in the form:
\begin{eqnarray}
\frac{da(t)}{dt}-a(t)\sqrt{{\kappa^2\over 3}\bigg(1+{\kappa^2\Phi(t)^2\over 8} \bigg)\bigg[{1\over 2}\bigg(\frac{d\Phi(t)}{dt}\bigg)^2+V(\Phi)\bigg]}=0\,,\label{A1}
\end{eqnarray}
 \begin{eqnarray}
\frac{d^2\Phi(t)}{dt^2}+3\frac{1}{a(t)}\frac{da(t)}{dt}\frac{d\Phi(t)}{dt}+{dV(\Phi)\over d\Phi}-{3\over 4} \bigg[\frac{1}{a(t)}\frac{da(t)}{dt}\bigg]^2{\Phi(t)\over (1+\kappa^2\Phi(t)^2/8)^2}=0\,,\label{A2}
\end{eqnarray}
The differential equation for $a(t)$ is of first order while for $\Phi(t)$ is second order. Given a potential $V(\Phi)$ we just need three initial conditions:
\begin{equation}
a(t=0)=a_i,\hspace{0.5cm} \Phi(t=0)=\Phi_i, \hspace{0.5cm} \frac{d\Phi(t=0)}{dt}=\dot{\Phi}_i
\end{equation}
We use:
\begin{equation}
V(\Phi)=v_0+\Lambda^4\Bigg(1-\frac{\Phi^2}{\sigma^2}\Bigg)^2 \,.
\end{equation}

For the Figures 2 (a), (b), (c) and (d), Figure 3 (a), (b), (c) and (d) we have used $a_i=1$, $\Phi_i=2.1m_{pl}=3.9\times 10^{43}$s$^{-1}$ and $\dot{\Phi}_i=0$, $\sigma = 1.0m_{pl}$, $v_0=0$ and $\Lambda=5\times 10^{-6}$. In Figure 4 (black line) we use the same parameters before with $v_0=1\times 10^{-28}m_{pl}^4$. We have used $\kappa^2=8\pi/m_{pl}^2$ with $m_{pl}=1.22\times 10^{19}$GeV$=1.86\times 10^{43}$s$^{-1}$. Thus all the time scales are in units of seconds.

For the Figure 3 (b) notice that the pressure (\ref{pphiElko}) contain a term $\dot{H}$, which requires an initial condition for $\dot{a}(t=0)$. To work around this issue we have used the expression (\ref{Hdot}) for $\dot{H}$ in the form:
\begin{equation}
\dot{H}=-{\kappa^2\over 2}\bigg(1+{\kappa^2\Phi(t)^2\over 8} \bigg)\bigg[ \bigg(\frac{d\Phi(t)}{dt}\bigg)^2-{1\over 2}\frac{1}{a(t)}\frac{da(t)}{dt}{\Phi(t) \over (1+\kappa^2\Phi(t)^2/8)^2}\frac{d\Phi(t)}{dt} \bigg]\,,\label{A4}
\end{equation}
into the corresponding term of (\ref{pphiElko}), thus the differential equation is of first order and the above initial conditions are sufficient to evaluate $\omega(t)$.


\begin{acknowledgements}
 SHP is grateful to CNPq - Conselho Nacional de Desenvolvimento Cient\'ifico e Tecnol\'ogico, Brazilian research agency, for financial support, grants number 304297/2015-1 and 400924/2016-1. TMG was partially supported by CAPES - Brazil.
\end{acknowledgements}

\twocolumngrid


\end{document}